\documentclass[prb,twocolumn,showpacs,showkeys,amssymb]{revtex4}

\usepackage{graphicx}
\usepackage{dcolumn}
\usepackage{bm}

\begin{document}


\title{Inelastic light scattering and the off-resonance approximation}
\author{Augusto Gonzalez}
\affiliation{Instituto de Cibernetica, Matematica
 y Fisica, Calle E 309, Vedado, Ciudad Habana, Cuba}
\author{Alain Delgado}
\affiliation{Centro de Aplicaciones Tecnologicas y Desarrollo Nuclear,
 Calle 30 No 502, Miramar, Ciudad Habana, Cuba}

\begin{abstract}
Inelastic (Raman) light scattering intensities for a 42-electron
quantum dot under off-resonance conditions and in different spin and angular momentum
channels are computed in order to test whether final collective states become the dominant
peaks in the process. The results of the calculations set a limit to the spread use of the
off-resonant approximation for the theoretical description of Raman processes.
\end{abstract}


\maketitle

\section{Introduction}

The spectroscopy of low-dimensional semiconductor structures has found in the inelastic 
(Raman) scattering of light a powerful tool \cite{ILS2,ILS3}. By means of this 
technique, a whole sector of the electronic excitation spectrum, not reachable from
direct light absorption because of Kohn theorem \cite{Kohn}, becomes the subject of 
research. Single-particle and collective states, spin- and charge excitations leave a
trace in the Raman spectra.

The interpretation of the results of a Raman experiment, on the other hand, is usually 
based on a simplified theoretical expression, the so called off-resonance approximation 
(ORA) \cite{ILS1}. The ORA states that, far away from resonance conditions, the Raman
scattering amplitude is dominated by collective states, i. e., by states carrying on
most of the strength of multipole operators.

Experiments in quantum wells and wires under extreme resonance, however, have revealed Raman peaks associated with single-particle excitations (SPE) \cite{exp1}. These peaks, absent in the ORA, are known to be related to taking a proper account of
the intermediate (virtual) states \cite{Sarma}. For higher
excitation energies, 40-50 meV above the band gap, a
resonant enhancement of Raman intensities for particular values of
the incident laser energy has been observed \cite{Danan}.
This resonant effect is clearly not described by the ORA, and has been ascribed to the
existence of incoming and outgoing resonances in the intermediate states.

The relevant experimental facts concerning electronic Raman
scattering in quantum dots can be found in Ref. [\onlinecite{Lockwood}],
and specially in paper [\onlinecite{Heitmann}]. As the incident laser energy moves from extreme resonance to 40 meV above it, the observed Raman spectrum evolves
from a SPE-dominated one to a spectrum dominated by collective
excitations. The positions of collective excitations for the dots
studied in Ref. [\onlinecite{Heitmann}] have been computed in Ref.
[\onlinecite{Lipparini}] by means of ORA expressions. This is a pretty example in which the results are partially valid in spite of the fact that the physical conditions for the application of the ORA are not.

In the present paper, we perform a critical examination of the validity of the ORA by
computing Raman scattering amplitudes for a 42-electron quantum dot. The position of Raman peaks are then compared with the position of collective states, obtained from the evaluation of multipole operators. Let us stress that the typical dots used in the experiments reviewed in [\onlinecite{Lockwood}] contain around 200 electrons. We hope that our 42-electron dot captures the salient features of these larger systems. Notice that, to the best of our knowledge, Raman calculations have been performed only for up to 12-electron dots whithin a restricted Hartree-Fock scheme \cite{calc}.

The plan of the paper is as follows. In the next two sections we include, for
completeness, a summary of formulas for Raman scattering and a deduction of the ORA.
Detailed calculations of Raman amplitudes and comparison with ORA results are presented 
in Sec. \ref{sec4}. Concluding remarks are given in Sec. \ref{sec5}.

\section{Raman scattering by electronic excitations}
\label{sec2}

Formulas for Raman scattering are deduced from ordinary second-order perturbation 
theory \cite{text}. For the sake of completeness, we provide a summary of 
these formulas in the present section.

The starting point is the total Hamiltonian:

\begin{equation}
{\cal H}={\cal H}_0^{(e)}+{\cal H}_0^{(rad)}+{\cal H}_I.
\end{equation}

\noindent
${\cal H}_0^{(e)}$ describes the electronic subsystem in the semiconductor structure. 
Coulomb interactions among electrons are included in it, but no electron-phonon 
interactions are to be considered. This means that excitation energies should be below 
the threshold for the creation of a longitudinal optical (Frohlich) phonon (around 30 
meV for GaAs), which is the main electron-phonon interaction mechanism in these 
structures \cite{Shah}. ${\cal H}_0^{(rad)}$ describes the free (quantized) photon
field, and ${\cal H}_I$ is the electron-radiation interaction Hamiltonian:

\begin{equation}
{\cal H}_I=\frac{1}{2 m_0}\sum_i \left\{ 2 e \vec A(\vec r_i) \cdot \vec p_i +
 e^2 |\vec A(\vec r_i)|^2 \right\}.
\end{equation}

\noindent
$m_0$ is the electron mass in vacuum. ${\cal H}_I$ contains a linear and a quadratic in
$\vec A$ terms. In second quantization formalism, these terms take the form:

\begin{widetext}
\begin{eqnarray}
{\cal H}_I^{(A)}=\frac{e}{m_0}\sum_k\sum_{\alpha,\beta}\left\{
 \langle\alpha|e^{i\vec k\cdot\vec r}\vec\varepsilon_k\cdot\vec p |\beta\rangle~
 a_k e^{\dagger}_{\alpha} e_{\beta} +
 \langle\alpha|e^{-i\vec k\cdot\vec r}\vec\varepsilon_k^*\cdot\vec p |\beta\rangle~
 a_k^{\dagger} e^{\dagger}_{\alpha} e_{\beta} \right\},
\label{eq3}
\end{eqnarray}

\begin{eqnarray}
{\cal H}_I^{(A^2)}&=&\frac{e^2}{2 m_0}\sum_{k,k'}\sum_{\alpha,\beta}\left\{
 \langle\alpha|e^{i(\vec k+\vec k')\cdot\vec r}\vec\varepsilon_k\cdot\vec 
  \varepsilon_{k'} |\beta\rangle~ a_k a_{k'} e^{\dagger}_{\alpha} e_{\beta} +
 \langle\alpha|e^{i(\vec k-\vec k')\cdot\vec r}\vec\varepsilon_k\cdot\vec 
  \varepsilon_{k'}^* |\beta\rangle~ a_k a_{k'}^{\dagger} e^{\dagger}_{\alpha} e_{\beta} 
 \right.\nonumber\\
 &+& \left. \langle\alpha|e^{-i(\vec k-\vec k')\cdot\vec r}\vec\varepsilon^*_k\cdot\vec 
  \varepsilon_{k'} |\beta\rangle~ a_k^{\dagger} a_{k'} e^{\dagger}_{\alpha} e_{\beta} +
 \langle\alpha|e^{-i(\vec k+\vec k')\cdot\vec r}\vec\varepsilon_k^*\cdot\vec 
  \varepsilon_{k'}^* |\beta\rangle~ a_k^\dagger a_{k'}^\dagger e^\dagger_\alpha e_{\beta} 
 \right\},
\label{eq4}
\end{eqnarray}
\end{widetext}

\noindent
where $a^\dagger, ~ a$ ($e^\dagger,~e$) are photon (electron) creation and annihilation 
operators, and $\vec\varepsilon$ is the light polarization vector.

The total wavefunction, $|\Psi\rangle$, is taken as a product of free photon and 
electronic functions. The initial state is made up from the ground state of the
electronic subsystem and an initial photon state. On the other hand, in the final state
the electronic subsystem is in an excited state, and there is a photon which
energy (and wavevector) changed from $h\nu_i$ to $h\nu_f$. The transition amplitude, to second order in the photon field, has contributions from ${\cal H}_I^{(A)}$ and
${\cal H}_I^{(A^2)}$:

\begin{widetext}
\begin{eqnarray}
C_{fi}=\frac{e^{i(h\nu_f+E_f-h\nu_i-E_i)t/\hbar}-1}{h\nu_f+E_f-h\nu_i-E_i}
 \left\{-\langle \Psi_f|{\cal H}_I^{(A^2)}|\Psi_i \rangle +
 \sum_{\Psi_{int}}\frac{\langle \Psi_f|{\cal H}_I^{(A)}|\Psi_{int}\rangle
 \langle \Psi_{int}|{\cal H}_I^{(A)}|\Psi_i \rangle}
 {\hbar\omega_{int}}\right\}.
\end{eqnarray}
\end{widetext}

\noindent
In the last equation, $E$ refers to the energy of the electronic system.
$\hbar\omega_{int}=E_{int}-(E_i+h\nu_i)$ is the difference between the (total)
energies of the intermediate
and initial states. The sum in brackets runs over intermediate states giving nonzero
matrix elements of ${\cal H}_I^{(A)}$. In the experiments, the incident photon energy
is close to the band gap energy, thus we restrict the sum to intermediate states with
an additional electron-hole pair, and will consider only the so called resonant
contribution. It may be shown (and numerical calculations support this statement)
that this is the main contribution to the sum. Taking into account Eqs.
(\ref{eq3},\ref{eq4}), and computing explicitly the matrix elements of photon
operators, one gets:

\begin{widetext}
\begin{eqnarray}
C_{fi}&=&\frac{e^2\sqrt{N_i(N_f+1)} A_0^2}{m_0}
 \left\{-\vec\varepsilon_i\cdot\vec\varepsilon_f \left\langle
  f\left|\sum_{\alpha,\beta}
  \langle\alpha|e^{i(\vec k_i-\vec k_f)\cdot\vec r}|\beta\rangle
  e^\dagger_\alpha e_\beta\right|i\right\rangle\right.\nonumber\\
 &+&\left.\frac{P^2}{m_0}\sum_{int}\frac{\langle f|H_{e-r}^+|int\rangle
  \langle int|H_{e-r}^-|i \rangle}{E_{int}-E_i-h\nu_i+i\Gamma_{int}}
 \right\}\frac{e^{i\omega_{fi}t}-1}{\hbar\omega_{fi}}.
\label{eq6}
\end{eqnarray}
\end{widetext}

\noindent
In Eq. (\ref{eq6}), $N_i$ ($N_f$) represents the number of photons in the initial
(scattered) beam. We will take $N_f=0$. Additionally, we have introduced the width,
$\Gamma_{int}$, of the level $E_{int}$. $P$ is the interband matrix element of $p$
(consequently, in the next formulas the momentum operator is a dimensionless
magnitude), and the $H_{e-r}^\pm$ are given by:

\begin{eqnarray}
H_{e-r}^-=\sum_{\alpha,\bar\beta}\langle \alpha|e^{i\vec
 k_i\cdot\vec r} \vec\varepsilon_i\cdot\vec p|\bar\beta\rangle ~
 e_\alpha^\dagger h^\dagger_{\bar\beta},
\end{eqnarray}

\begin{eqnarray}
H_{e-r}^+=\sum_{\bar\alpha,\beta}\langle \bar\alpha|e^{-i\vec
 k_f\cdot\vec r} \vec\varepsilon_f^*\cdot\vec p|\beta\rangle ~
 h_{\bar\alpha}e_\beta,
\end{eqnarray}

\noindent
where we have introduced hole operators in the valence band.

Eq. (\ref{eq6}) is the main expression to be used in the following sections. In Sec.
\ref{sec4}, we present results of Raman scattering in a quantum dot based on
this formula. Notice that the numerical evaluation of Eq. (\ref{eq6}) requires obtaining reliable approximations to the electronic wavefunctions $|i\rangle$, $|f\rangle$, $|int\rangle$, and the eigenvalues $E_i$, $E_f$, $E_{int}$. In the paper, we use RPA-like wavefunctions, obtained with the help of methods of Nuclear Physics \cite{nuclear}, which are especially intended for finite systems. Details of the procedure can be found in Refs. [\onlinecite{nuestro,DGL}].

\section{The off-resonance approximation}
\label{sec3}

In this section, the ORA is deduced from Eq. (\ref{eq6}) in the limit when the incident laser energy is far from resonance.

Let us consider the two terms in curly brackets in Eq. (\ref{eq6}). They are
dimensionless quantities. The first term comes from ${\cal H}_I^{(A^2)}$. It has
the form of a structure function. Greek letters $\alpha,~\beta$ in Eq. (\ref{eq6}) 
denote a complete set of (conduction-band) one-electron states. In this section, we use a different notation for these states, and will write $\alpha\uparrow$ or $\alpha\downarrow$, in which $\alpha$ labels the orbital wavefunction, and the spin projection is explicitly indicated. The first term is thus rewritten:

\begin{eqnarray}
T_1=-\vec\varepsilon_i\cdot\vec\varepsilon_f
 \sum_{\alpha,\beta}\langle\alpha|e^{i(\vec k_i-\vec k_f)\cdot\vec r}|\beta\rangle
 \langle f|e^\dagger_{\alpha\uparrow} e_{\beta\uparrow}+
 e^\dagger_{\alpha\downarrow} e_{\beta\downarrow}|i\rangle.\nonumber\\
\label{eq9}
\end{eqnarray}

\noindent
On the other hand, the sum in Eq. (\ref{eq6}) has in front of it a big coefficient, $P^2/m_0$. In GaAs, for example, this magnitude is roughly 11.35 eV \cite{Bastard}.
The denominator, on his side, contains the difference $E_{int}-E_i$, which is always greater than the effective band gap, $E_{gap}$. The sum takes big values when the denominator is near zero, $E_{int}-E_i\approx h\nu_i$ (resonance). To obtain the ORA, however, we shall assume that the photon energy is far away from resonance. The energy denominator is so big that we can neglect its variation with $E_{int}$. Then, we write
$E_{int}-E_i-h\nu_i\approx E_{gap}-h\nu_i$, and use the completeness relation for the
intermediate states. In this way, we get for the second term (the sum):

\begin{equation}
T_2=\frac{P^2/m_0}{E_{gap}-h\nu_i}\langle f|H^+_{e-r}H^-_{e-r}|i\rangle.
\label{eq10}
\end{equation}

Next, we notice that $|i\rangle$ and $|f\rangle$ do not contain hole states. Thus,
the hole operators in Eq. (\ref{eq10}) should contract to a Kronecker delta
function. We took for the hole a complete basis of states labeled by the orbital
index, $\bar\gamma$, and its total (band) angular momentum projection, $m_j$. $T_2$
takes the form:

\begin{equation}
T_2=\sum_{\alpha,\alpha'}\{[\uparrow\uparrow]+[\downarrow\downarrow]+
[\uparrow\downarrow]+[\downarrow\uparrow]\},
\end{equation}

\noindent
where we have used the abbreviated notation:

\begin{widetext}
\begin{eqnarray}
[\uparrow\uparrow]=\frac{P^2/m_0}{E_{gap}-h\nu_i}
 \sum_{m_j,\bar\gamma} \langle \alpha\uparrow|e^{i\vec q_i\cdot\vec r}
 \vec\varepsilon_i\cdot\vec p |\bar\gamma,m_j \rangle
 \langle \bar\gamma,m_j|e^{-i\vec q_f\cdot\vec r}
 \vec\varepsilon_f^*\cdot\vec p |\alpha'\uparrow \rangle
 \langle f|e_{\alpha'\uparrow}e^\dagger_{\alpha\uparrow}|i\rangle,
\end{eqnarray}
\end{widetext}

\noindent
etc. Matrix elements of factors like $\vec\varepsilon\cdot\vec p$ depend only
on the electron spin projection and the hole index $m_j$. They are computed according
to Table \ref{tab1}, in which the components $\varepsilon_\pm$ are defined
as $\varepsilon_\pm=\mp (\varepsilon_x\mp i\varepsilon_y)/\sqrt{2}$. Finally,
we use completeness of the hole orbital wave functions. The result is as follows:

\begin{widetext}
\begin{eqnarray}
T_2= &-&\frac{P^2/m_0}{E_{gap}-h\nu_i}\sum_{\alpha,\alpha'}
 \langle\alpha| e^{i(\vec q_i-\vec q_f)\cdot \vec r}|\alpha'\rangle
 \left\langle f\left|
 \left\{\frac{2}{3} (\vec \varepsilon_i\cdot\vec\varepsilon_f)~
 [e^\dagger_{\alpha\uparrow} e_{\alpha'\uparrow}+
 e^\dagger_{\alpha\downarrow} e_{\alpha'\downarrow} ]
 \right.\right.\right.\nonumber\\
&+&\frac{i}{3} (\vec \varepsilon_i\times\vec\varepsilon_f)\cdot \hat z~
 [e^\dagger_{\alpha\uparrow} e_{\alpha'\uparrow}-
 e^\dagger_{\alpha\downarrow} e_{\alpha'\downarrow}]\nonumber\\
&+& \left.\left.\left.
\frac{i}{3} (\vec \varepsilon_i\times\vec\varepsilon_f)\cdot
(\hat x+i\hat y)~  e^\dagger_{\alpha\uparrow} e_{\alpha'\downarrow}
+\frac{i}{3} (\vec \varepsilon_i\times\vec\varepsilon_f)\cdot
(\hat x-i\hat y)~ e^\dagger_{\alpha\downarrow} e_{\alpha'\uparrow}
    \right\}\right|i\right\rangle.
\label{eq13}
\end{eqnarray}
\end{widetext}

In the ORA, the Raman amplitude $C_{fi}$ is proportional to $T_1+T_2$, where $T_1$ and $T_2$ are given by Eqs. (\ref{eq9}) and (\ref{eq13}). Notice that, in this approximation, the intermediate states play no role in the Raman amplitude.

For simplicity, in what follows we shall consider the quasi-twodimensional motion of electrons. An external harmonic potential, with characteristic energy $\hbar\omega_0$,
confines the electron motion to a region in the plane: the quantum dot.

In order to realize that only collective states contribute to $T_1+T_2$, let us expand the exponential $e^{i(\vec q_i-\vec q_f)\cdot \vec r}$ in Eq. (\ref{eq13}). The main term of this expansion, for example, leads to a factor:

\begin{eqnarray}
D^0_{fi}=\left\langle f\left| \sum_{\alpha,\alpha'}
 d^0_{\alpha,\alpha'}[ e^\dagger_{\alpha\uparrow} e_{\alpha'\uparrow}+
 e^\dagger_{\alpha\downarrow} e_{\alpha'\downarrow} ]
 \right|i\right\rangle.
\end{eqnarray}

\noindent
i.e., the matrix element of the monopole operator. The coefficients $d^l_{\alpha\gamma}$ are defined according to \cite{nuestro}:

\begin{eqnarray}
d^l_{\alpha\gamma}&=&\langle \alpha|\rho^{|l|} e^{i l \theta}
 |\gamma\rangle;~~l\ne 0,\nonumber \\
&=&\langle \alpha|\rho^2|\gamma\rangle;~~l=0,
\end{eqnarray}

\noindent
where $\rho$ and $\theta$ are polar coordinates in the plane.

Final states, $|f\rangle$, for which $D^0_{fi}$ is significantly different from zero are called collective charge monopole excitations of the initial (ground) state $|i\rangle$. The exact quantitative meaning of this statement comes from the fact that $D^0$ satisfies the energy-weighted sum rule \cite{nuclear}:

\begin{equation}
\sum_f (E_f-E_i) |D^0_{fi}|^2=\frac{2\hbar^2}{m_e}
 \sum_{\boldsymbol{\lambda}\le\mu_F} \langle \boldsymbol{\lambda}|\rho^2|
 \boldsymbol{\lambda} \rangle,
\end{equation}

\noindent
where $\mu_F$ is the electron Fermi level. A state is (conventionally) said
to be collective if its contribution to the sum rule is greater than 5 \% of the total.
Similar criteria can be formulated for charge multipole states ($|l|>0$),
\cite{nuestro} or for spin excited states (involving or not
spin reversal with respect to the ground state). The latter are
related to the last three terms of Eq. (\ref{eq13}), proportional to
$\vec \varepsilon_i\times\vec\varepsilon_f$.

We may conclude this section by saying that, in the ORA, only charge and spin collective states contribute to the Raman amplitude $C_{fi}$, which is proportional to $T_1+T_2$.

\section{Numerical results}
\label{sec4}

Below, we present results of calculations of Raman scattering in a
quantum dot with 42 electrons. Details of the numerical procedure can be found
elsewhere \cite{nuestro,DGL}. We use the model parameters (effective masses,
dielectric constant, Kohn-Luttinger parameters, etc) of Ref. [\onlinecite{DGL}].

We shall compute the expression in curly brackets of Eq. (\ref{eq6}). The first
term, proportional to $\vec\varepsilon_i\cdot\vec\varepsilon_f$, shows to be
negligible in all our computations. Thus, it will not be included in the following.
We will also exclude the big factor, $P^2/m_0$, and will show results for the
smoothed function

\begin{widetext}
\begin{equation}
I(E)=\sum_f\left |\sum_{int}\frac{\langle f|H_{e-r}^+|int\rangle
  \langle int|H_{e-r}^-|i \rangle}{E_{int}-E_i-h\nu_i+i\Gamma_{int}}
  \right |^2 \frac{\Gamma_f/\pi}{(E-E_f)^2+\Gamma_f^2}.
\label{eq17}
\end{equation}
\end{widetext}

\noindent
$I$ is proportional to the differential cross section of the process.
The Lorentzian in Eq. (\ref{eq17}) is an approximation to the delta function
coming from $|C_{fi}|^2$ for large values of $t$. The
phenomenological parameters, $\Gamma_{int},~\Gamma_f$ are fixed to 0.5 and
0.1 meV, respectively. The sum runs over intermediate states with excitation
energy below 30 meV. This means states for which $E_{int}-E_{int}^{(0)}<30$ meV,
where $E_{int}^{(0)}$ is the energy of the lowest intermediate state.
$E_{int}^{(0)}$ enters the definition of the qdot effective band gap:
$E_{gap}=E_{int}^{(0)}-E_i$.

To approach the ORA regime, the laser excitation
energy, $h\nu_i$, will range in an interval below $E_{gap}$.\cite{nota} It will be shown
that 30 meV below $E_{gap}$ is enough.

Calculations are performed in backscattering geometry, where both the
incident and scattered light form an angle of $20^\circ$ with the dot normal.

As ORA does not provide the correct relative intensities of
the collective states, comparison between numerical calculations from Eq. (\ref{eq17}) and the ORA expression, $T_1+T_2$, is done at a qualitative level. The position of collective states are obtained from the multipole operators. When $I$ shows to be dominated by collective states, we say that the ORA regime has been reached.

\begin{figure*}[ht]
\begin{center}
\includegraphics[width=.8\linewidth,angle=0]{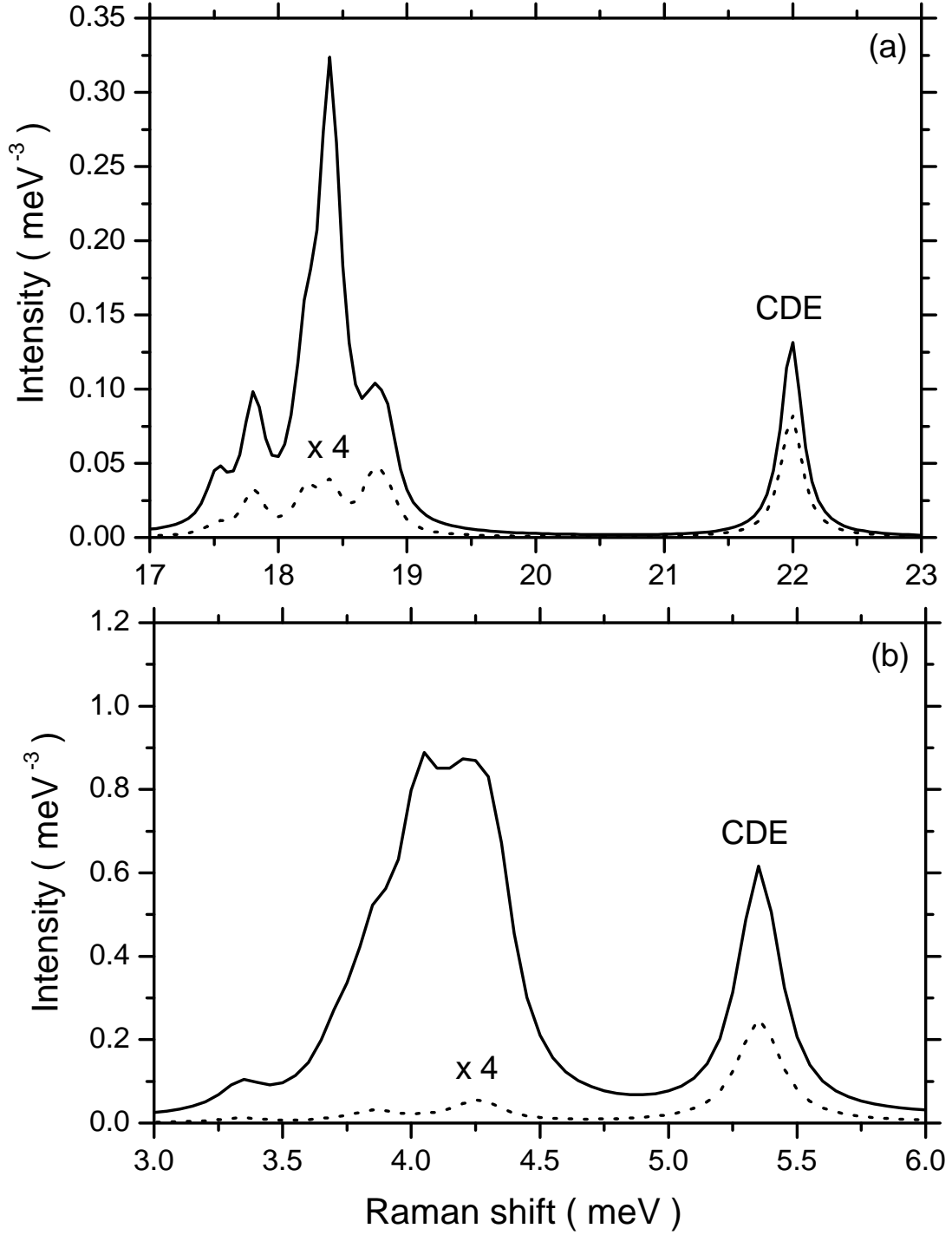}
\caption{\label{fig1} Raman intensities for charge monopole final states and polarized
geometry: (a) $\hbar\omega_0=12$ meV, (b) $\hbar\omega_0=3$ meV. Two curves are shown
corresponding to $h\nu_i=E_{gap}+5$ meV (solid line), and $h\nu_i=E_{gap}-30$ meV (dashed line).
This convention is used in the next two figures.}
\end{center}
\end{figure*}

We show in Fig. \ref{fig1} the Raman intensity, $I$, in the charge monopole
channel. This means that final electronic states in the scattering process are
monopole excitations ($\Delta L=0$) with the same total spin projection as the
initial ground state, $|i\rangle$. Computations are carried out in the so called
polarized Raman geometry, in which the initial and final light polarization
vectors are parallel. The $x$ axis represents the Raman shift, $E-E_i$.
The position of the collective state, labeled
CDE (charge density excitation), is explicitly signalized. Peaks corresponding to
single-particle excitations are positioned to the left of the CDE. Two curves are
drawn in each panel. The first one (shown for comparative purposes) is obtained with a laser excitation energy $h\nu_i=E_{gap}+5$ meV, and the second one with $h\nu_i=E_{gap}-30$ meV. A scaling factor is introduced for the sake of clarity.

We see a distinct tendency toward dominance of the CDE state for below band gap
excitation, an indication that the ORA regime is being reached. This tendency is
reinforced when the in-plane confinement, $\hbar\omega_0$, is lowered. The reason is simple. To compute $I$, we use intermediate states in the
interval [$E_{gap},E_{gap}+30$ meV]. When $\hbar\omega_0$ is decreased by a factor $\eta$, the number of states in this interval increases roughly by the same factor, and
the assumption made in the derivation of the ORA concerning completeness of the
set of intermediate states is better fulfilled. We show in Fig. \ref{fig1} curves for $\hbar\omega_0=12$ and 3 meV as examples. Let us stress
that the argument to restrict the sum over intermediate states to this energy interval is
more than technical, and is related to the sudden increase of $\Gamma_{int}$ for
larger excitation energies, as will be discussed below.

\begin{figure*}[h]
\begin{center}
\includegraphics[width=.8\linewidth,angle=-90]{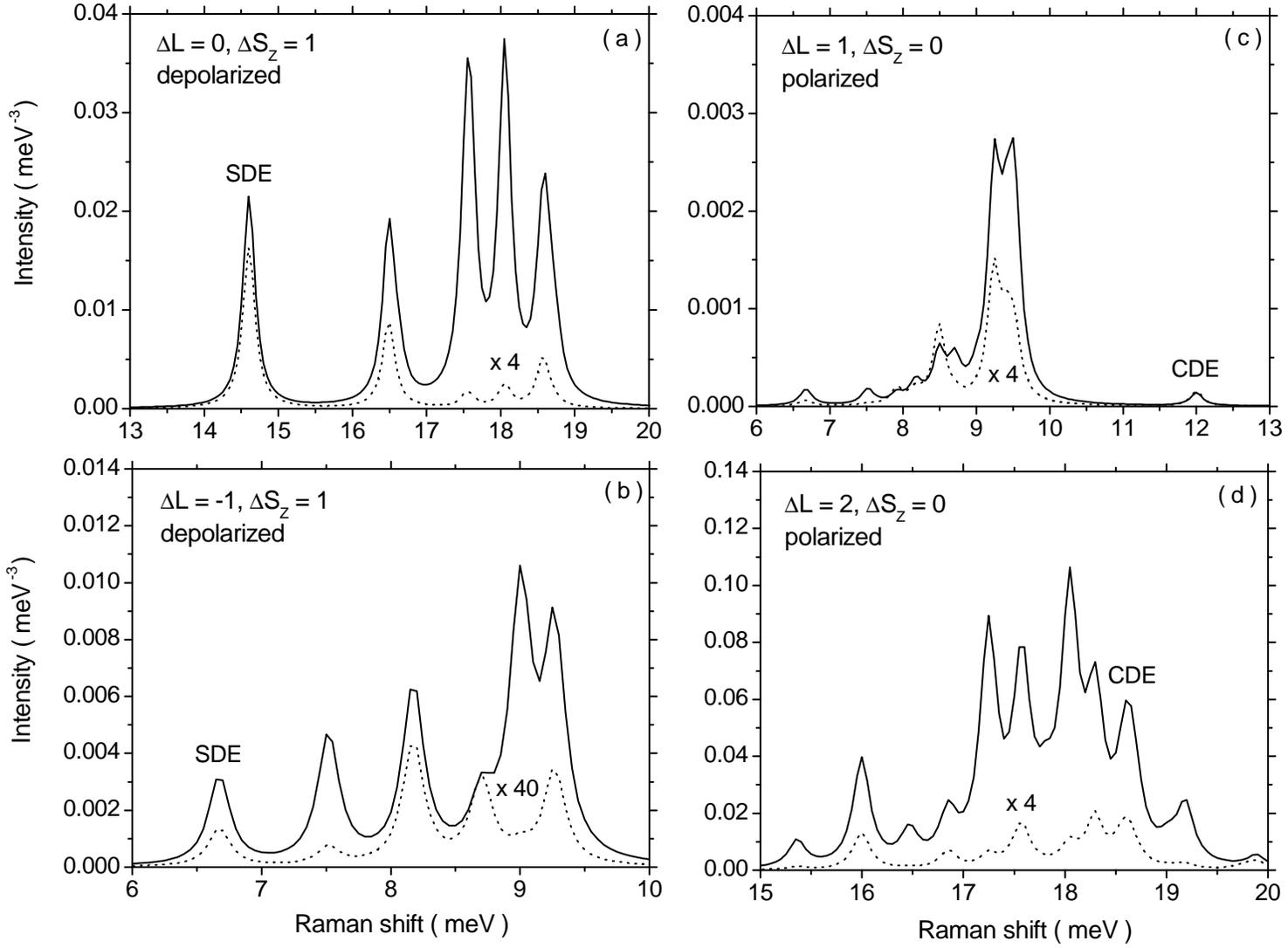}
\caption{\label{fig2} Raman intensities in different spin and angular momentum channels,
and for $\hbar\omega_0=$12 meV.}
\end{center}
\end{figure*}

\begin{figure*}[ht]
\begin{center}
\includegraphics[width=.8\linewidth,angle=-90]{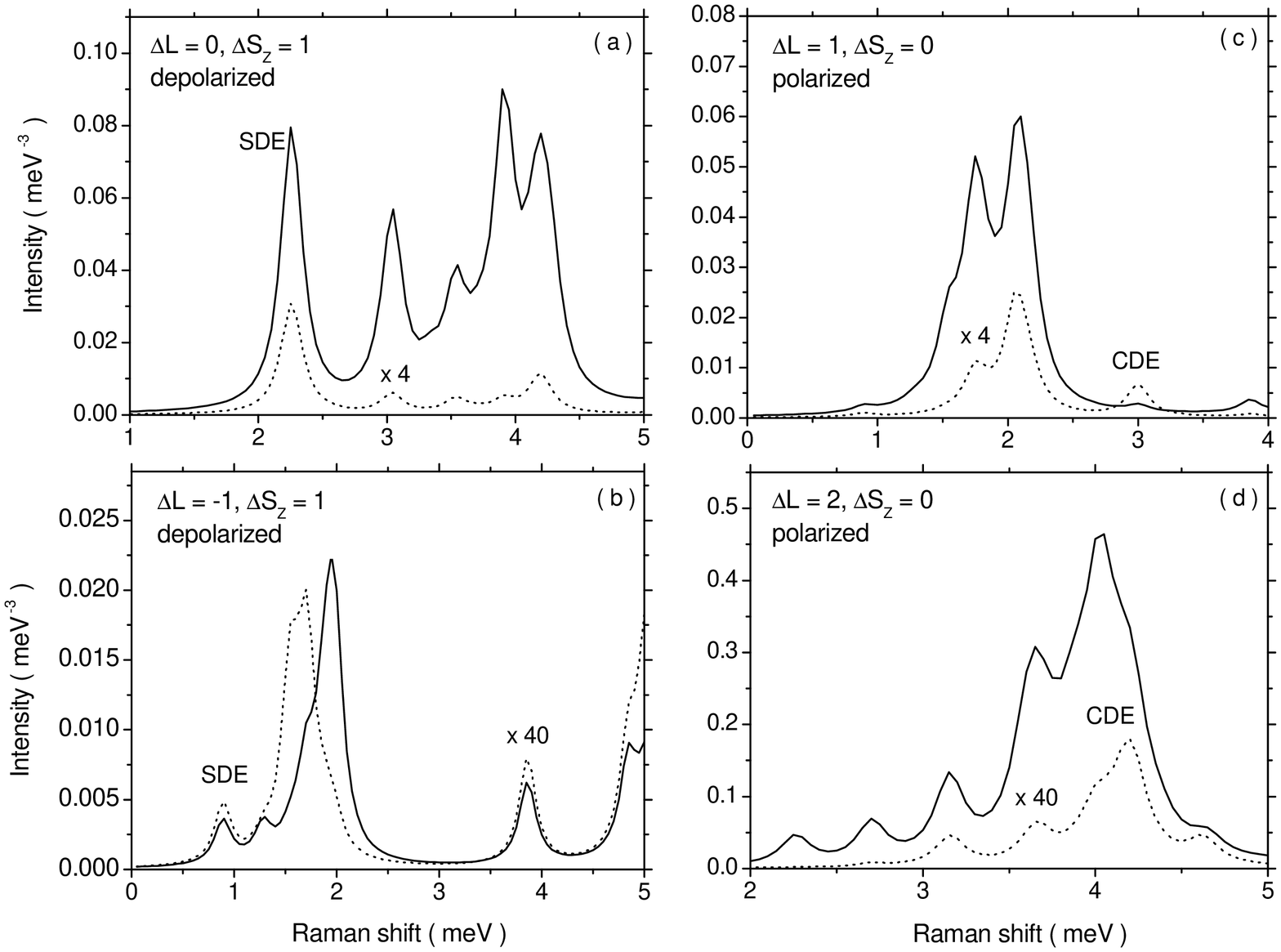}
\caption{\label{fig3} Same as in Fig. \ref{fig2}, but for $\hbar\omega_0$=3 meV.}
\end{center}
\end{figure*}

Figs. \ref{fig2} and \ref{fig3} show results of calculations in other channels.
The monopole spin-flip case, $\Delta L=0$, $\Delta S=1$, computed in the
depolarized Raman geometry, $\vec\varepsilon_f\cdot\vec\varepsilon_i=0$, is drawn
in the left upper corner. The collective state, labeled SDE (spin density
excitation), lies to the left of the single-particle peaks in this case. The
behavior when $h\nu_i$ is varied is similar to the charge monopole situation
discussed above. The SDE dominates the spectrum for below band gap excitation, and
this tendency is better observed for $\hbar\omega_0=3$ meV.

In the charge quadrupole channel, $\Delta L=2$, $\Delta S=0$, and
polarized Raman geometry, right lower panel, the dominance of the CDE is observed
only when $\hbar\omega_0$ is lowered to 3 meV, accompanied by a strong decrease
of the intensity.

Dipole final states, $\Delta L=\pm 1$, however, show a different behavior in which
collective states are never the dominant peaks in Raman scattering. The charge dipole
channel, $\Delta L=1$, $\Delta S=0$, computed in the polarized geometry, is shown in
the right upper panel. Notice that the CDE is not a significant peak even with
above band gap excitation. In the spin-flip case, left lower corner, the SDE state
is always a low peak. Notice, in addition, that in the latter case the intensity
experiences a strong decrease for below band gap excitation.

\section{Discussion}
\label{sec5}

We have computed Raman intensities from a 42-electron qdot in different angular
momentum and spin channels for below band gap excitation to show how and to what
extent the ORA regime is reached.

It was demonstrated that for (charge and spin)
monopole ($\Delta L=0$) final states, collective excitations rapidly become the
dominant excited modes and, already for laser energy 30 meV below the effective
band gap, are the main peaks in the Raman spectra.

In any other channel, however, the ORA may or may not play a relevant role. We
observed, for example, the dominance of the CDE state in the charge quadrupole
channel when $\hbar\omega_0$ is lowered, and thus the density of states in
the dot is increased. But we observed no range of $h\nu_i$ in which collective modes
become the leading dipole peaks.

The conclusion is that the ORA works very well for monopole states, but only
conditionally for quadrupole ones, and very badly for dipole states. The number of
electrons in the dot or the strength of the external confining potential may dictate
the way in which the ORA regime is reached by varying the density of energy levels
in the dot. Experiments in quantum dots for below band gap excitation could confirm this conclusion.

More generally speaking, we suggest caution when an experimental Raman peak is to be interpreted as a collective excitation, on the basis of ORA arguments, in other systems like quantum wires or wells. Theoretical calculations should support this kind of assignements.

Collective states could dominate a Raman spectrum in a regime very different to that one
leading to the ORA. Let us recall again the results presented in
paper [\onlinecite{Heitmann}] for laser excitation
energies 40 - 50 meV above the effective band gap. In this energy interval there are plenty of intermediate states, thus the experiment deals with a resonant regime.
A correct theoretical description should, probably, take account of the dependence
of $\Gamma_{int}$ on $E_{int}$.\cite{DGL} On qualitative grounds, one can expect a sudden
increase of $\Gamma_{int}$ when $E_{int}$ is over 40 meV above the effective band gap. There could be, however, special intermediate states (which can be thought of as ``plasmon
plus exciton'' states) for which $\Gamma_{int}$ may preserve relatively small values.
The physics of Raman scattering in this regime is based, in our opinion, on the
existence of these intermediate state resonances, and has no relation with the ORA.
This interesting problem deserves more theoretical and experimental work on it.

A second example of collective states dominating Raman scattering, and in which the ORA has no relevance, is given in Ref.
[\onlinecite{Pinczuk}], where the peak related to the electronic cyclotron resonance
in a quantum well undergoes an abrupt increase when it moves through the position of a
luminescence peak (this is achieved by varying $h\nu_i$). The theoretical description,
we guess, should include the enhancement of Raman intensity due to the background
luminescence (The $N_f$ variable, which was set to zero in Eq. (\ref{eq6})) and a
consistent calculation of luminescence (including master equations for the population
of the intermediate states). This situation is, again, far from the simple assumptions
leading to the ORA.

Our final conclusion is, thus, that there are still many interesting experimental and
theoretical aspects of Raman scattering in semiconductor quantum structures waiting for
elucidation. Simple recipes such as the ORA give a first insight into the physics, but
more elaborated theoretical constructions are needed for a complete understanding of the
process.

\begin{table*}[h]
\begin{tabular}{|c||c|c|c|c|}
\hline
$\sigma \backslash ~m_j $& 3/2 & 1/2 & -1/2 & -3/2 \\
\hline\hline 1/2 & $\varepsilon _{+}$ &
$\sqrt{2/3}\;\varepsilon_{z}$ & $\sqrt{1/3}\;\varepsilon_{-}$ & 0 \\
\hline -1/2 & 0 & $\sqrt{1/3}\;\varepsilon _{+}$ & $\sqrt{2/3}\;
 \varepsilon_{z}$ & $\varepsilon_{-}$ \\
\hline
\end{tabular}
\caption{The matrix elements $\langle \sigma|\vec\varepsilon\cdot\vec p|m_j\rangle/i$.}
\label{tab1}
\end{table*}

\begin{thebibliography}{99}
\bibitem{nota} In most of the experiments, Raman spectra are taken with
 above band gap excitation. We stress that, excitation below band gap is
 not only useful to test the ORA, but also to clearly identify Raman peaks
 \cite{DGL}.
\end{thebibliography}
\end{document}